\documentclass[conference, final, letterpaper]{IEEEtran}
\usepackage{cite}
\usepackage{amsmath,amssymb,amsfonts}
\usepackage{algorithmic}
\usepackage{graphicx}
\usepackage{textcomp}
\usepackage{xcolor}
\usepackage{mathptmx}

\usepackage[multiple]{footmisc}
\usepackage{tikz}
\usepackage{pgfplots}
\pgfplotsset{compat=1.17}
\usetikzlibrary{decorations.pathreplacing, positioning, patterns}
\usepgfplotslibrary{statistics, groupplots}
\DeclareMathOperator{\E}{E} 

\newcommand\copyrighttext{%
  \footnotesize \textcopyright~2021 IEEE. Personal use of this material is permitted. Permission from IEEE must be obtained for all other uses, in any current or future media, including reprinting/republishing this material for advertising or promotional purposes, creating new collective works, for resale or redistribution to servers or lists, or reuse of any copyrighted component of this work in other works.}
\newcommand\copyrightnotice{%
\begin{tikzpicture}[remember picture,overlay]
\node[anchor=south,yshift=10pt] at (current page.south) {\fbox{\parbox{\dimexpr\textwidth-\fboxsep-\fboxrule\relax}{\copyrighttext}}};
\end{tikzpicture}%
}

\begin{document}
\title{Avoiding normalization uncertainties in deep learning architectures for end-to-end communication\\
}

\author{\IEEEauthorblockN{Simon Bos, Evgenii Vinogradov, Sofie Pollin}
\IEEEauthorblockA{\textit{Department of Electrical Engineering, KU Leuven, Belgium} \\
Email: \{simon.bos, evgenii.vinogradov, sofie.pollin\}@kuleuven.be}
}

\maketitle
\copyrightnotice

\begin{abstract}
Recently, deep learning is considered to optimize the end-to-end performance of digital communication systems. The promise of learning a digital communication scheme from data is attractive, since this makes the scheme adaptable and precisely tunable to many scenarios and channel models. In this paper, we analyse a widely used neural network architecture and show that the training of the end-to-end architecture suffers from normalization errors introduced by an average power constraint. To solve this issue, we propose a modified architecture: shifting the batch slicing after the normalization layer. This approach meets the normalization constraints better, especially in the case of small batch sizes. Finally, we experimentally demonstrate that our modified architecture leads to significantly improved performance of trained models, even for large batch sizes where normalization constraints are more easily met.
\end{abstract}


\section{Introduction}
Digital communication is considered as a complex and mature engineering field in which modeling and tractable analytical models have shown to enable incredible solutions. Currently, digital communication systems are partitioned in independent blocks which are individually optimized for their specific tasks (e.g., source \& channel encoding or modulation) This makes the digital communication systems of today versatile and controllable, but could lead to a suboptimal end-to-end performance. Additionally, most signal processing algorithms are optimized for tractable analytical (channel) models while relying on some assumptions. The performance of the system could be lower under practical conditions. The application of deep learning to an end-to-end digital communication system shows potential to solve these issues~\cite{oshea2017introduction}.

\begin{figure}[t!]
    \centering
    \resizebox{\columnwidth}{!}{

\begin{tikzpicture}
\tikzstyle{hbox}=[fill=white, draw=black, shape=rectangle, minimum height=30pt, minimum width=60pt, align=center]
\tikzstyle{pointer}=[->]
\tikzstyle{pointer text}=[above]
		\node (in) {$i$};
		\node (tx) [style=hbox, right=15pt of in] {Transmitter\\$f_T: I \to \mathbb{C}$};
		\node (ch) [style=hbox, right=20 pt of tx] {Channel\\$p(y|x_i)$};
		\node (rx) [style=hbox, right=20pt of ch] {Receiver\\$f_R: \mathbb{C} \to I$};
		\node (out) [right=15pt of rx] {$\hat{i}$};
		\draw [style=pointer] (in.east) to (tx.west);
		\draw [style=pointer] (tx.east) to node[pointer text] {$x_i$} (ch.west);
		\draw [style=pointer] (ch.east) to node[pointer text] {$y$} (rx.west);
		\draw [style=pointer] (rx.east) to (out.west);
\end{tikzpicture}}
    \caption{General end-to-end digital communication system architecture.}
    \label{fig:arch-general}
\end{figure}
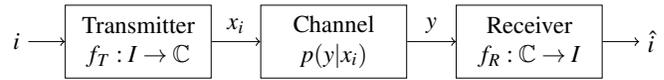

The idea of using deep learning for end-to-end digital communication was first proposed in~\cite{oshea2017introduction} and has been shown to have a competitive performance on additive white Gaussian noise (AWGN) and Rayleigh fading channels. Their proposed general end-to-end system architecture is outlined in Fig.~\ref{fig:arch-general} (a more detailed architecture is found in Fig.~\ref{fig:arch-detailed}). The aim of this system is to transmit a number of bits $K$ over the channel which the receiver can reconstruct.

Every permutation of $K$ bits is represented by a corresponding message $m_i,~\forall i \in I = \{1,2, \ldots ,M\}$ with $M=2^K$. The transmitter applies the function\footnote{Complex numbers are represented by $\mathbb{R}^2$ in most implementations, due to unsupported complex operations in deep learning frameworks.}\footnote{In~\cite{oshea2017introduction}, an extension to multiple channel uses is considered (i.e. $\mathbb{C}^n$).} $f_T: I \to \mathbb{C}$ to the message identified by $i$ to generate a transmitted signal $x_i$. Values of $x_i$ are usually constrained by the hardware. Next, the channel is described by its conditional probability density function $p(y|x_i)$, with $y$ the received signal. The receiver applies the function $f_R: \mathbb{C} \to I$ to $y$ to give an estimate $\hat{i}$ for the message that was transmitted.

The transmitter and receiver functions $f_T$ and $f_R$ are represented by a (deep) neural network (hardware constraints are represented by a normalization layer). This system is trained in a supervised fashion by uniformly sampling a batch of message identifiers, passing it through the system, computing the estimation error and updating the weights of both networks using backpropagation to obtain a minimal end-to-end estimation error. In this approach, performing backpropagation over the channel requires a differentiable model of the channel in which the system will be deployed. This model should include all possible real-world effects; otherwise the end-to-end performance is poor~\cite{dorner2018deep}. However, in practice only simplified channel models are available. Therefore, this approach cannot fully unlock the potential of deep learning.


Several works tackled the limitation of the original approach: \cite{dorner2018deep} investigated over-the-air fine-tuning, \cite{oshea2019approximating,ye2018channel} applied generative adversarial network to create a differentiable channel model and \cite{aoudia2019model} proposed a method based on policy gradients.
A disadvantage of the neural networks used in \cite{oshea2017introduction,dorner2018deep,oshea2019approximating,ye2018channel,aoudia2019model} is that only an approximation of the normalization constraints is used while training the networks. As we show in Sections~\ref{sec:proposal}~and~\ref{sec:experiments}, this leads to suboptimal trained models, even for larger batch sizes where the normalization approximation is improved.

In this article, we propose a modified version of the end-to-end architecture presented in \cite{oshea2017introduction} to better adhere to the normalization constraints; especially in the case of small batch sizes. The rest of this paper is organized as follows. In Section~\ref{sec:e2edetailed}, a more detailed version of Fig.~\ref{fig:arch-general} is explained. In Section~\ref{sec:proposal}, the proposed modified architecture is outlined together with the rationale. In Section~\ref{sec:experiments}, a comparison of performance between system architectures is conducted and the conclusions are drawn in Section~\ref{sec:conclusion}.

\section{Detailed end-to-end communication system}
\label{sec:e2edetailed}

\begin{figure}[t!]
\centering
\resizebox{\columnwidth}{!}{

\begin{tikzpicture}
\tikzstyle{none}=[]
\tikzstyle{hbox}=[fill=white, draw=black, shape=rectangle, minimum height=90pt, minimum width=25pt, align=center]
\tikzstyle{vbox}=[fill=white, draw=black, shape=rectangle, minimum height=25pt, minimum width=90pt, align=center, rotate=90, anchor=north]
\tikzstyle{pointer}=[->]
\tikzstyle{brace}=[thick, decoration={brace, mirror, raise=0.2cm}, decorate]
\tikzstyle{brace text}=[below=8pt]
\tikzstyle{brace up}=[decoration={brace, raise=0.2cm}, decorate]
\tikzstyle{brace up text}=[above=8pt]
\tikzstyle{pointer text}=[above]
		\node (tx1) [style=hbox] {$m_1$ \\ $\vdots$ \\ $m_i$ \\ $\vdots$ \\ $m_M$};
		\node (tx2) [right=3pt of tx1.east, style=vbox] {Encoding (one-hot)};
		\node (tx3) [right=3pt of tx2.south, style=vbox, fill=lightgray] {Index slicing};
		\node(in) [above=8pt of tx3.east] {$i$};

		\node (tx4) [right=15pt of tx3.south, style=vbox] {Dense Layers};
		\node (tx5) [right=3pt of tx4.south, style=vbox] {Normalization};

		\node (ch1) [right=25pt of tx5.south, style=vbox] {Noise};

		\node (rx1) [right=25pt of ch1.south, style=vbox] {Dense layers};
		\node (rx2) [right=2pt of rx1.south, style=vbox] {Dense layer \\ (softmax activation)};

		\node (out) [right=8pt of rx2.south] {$\hat{i}$};

		\draw [style=pointer] (in.south) to (tx3.east);
		\draw [style=pointer] (rx2.south) to (out.west);
		\draw [style=pointer] (tx3.south) to node[pointer text] {$s_i$} (tx4.north);
		\draw [style=pointer] (tx5.south) to node[pointer text] {$x_i$} (ch1.north);
		\draw [style=pointer] (ch1.south) to node[pointer text] {$y$} (rx1.north);

		\draw [style=brace up] (tx2.north east) to node[brace up text] {$s$} (tx2.south east);
		\draw [style=brace up] (rx2.north east) to node[brace up text] {$\hat{p}$} (rx2.south east);

		\draw [style=brace] (tx1.south west) to node[brace text] {Transmitter} (tx5.south west);
		\draw [style=brace] (ch1.north west) to node[brace text] {Channel} (ch1.south west);
		\draw [style=brace] (rx1.north west) to node[brace text] {Receiver} (rx2.south west);
\end{tikzpicture}}
\caption{Detailed end-to-end communication system over an AWGN channel, as proposed in~\cite{oshea2017introduction}.}
\label{fig:arch-detailed}
\end{figure}
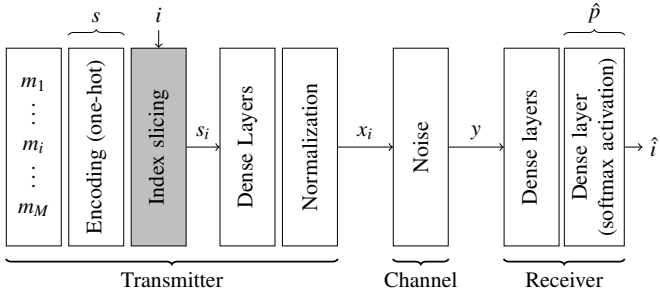

Fig.~\ref{fig:arch-detailed} shows the more detailed architecture for an AWGN channel presented in \cite{oshea2017introduction}. In its first part, the transmitter applies pre-processing to obtain $s_i$, the one-hot encoded version of message $m_i$ ($i$ is the input to the model). In the second part, the transmitter consists of a sequence of fully-connected/dense layers with a Rectified Linear Unit (ReLU) activation function for the first layers and a linear activation function for the last layer (which consists of 2 neurons representing $\mathbb{C}$). Afterwards, normalization is performed to ensure either a fixed power constraint $|x_i|^2 = P,~\forall i \in I$ or an average power constraint $\E\left[|x_i|^2,~\forall i \in I\right] = P$ for some fixed transmitter power~$P$.

The AWGN channel is represented by an additive noise layer with noise variance $\sigma^2$ derived from a fixed signal-to-noise ratio ($SNR$) defined as $SNR \triangleq P / \sigma^2$. Note that $SNR$ and $P$ can be considered implicit inputs of the end-to-end system.

The receiver consists of a sequence of fully-connected layers with a ReLU activation function for the first layers, but a softmax activation function at the last layer. The output of this layer $\hat{p}$ is a probability vector over all messages, and the index with the highest probability is taken as $\hat{i}$.

The end-to-end system is trained in a supervised fashion by firstly uniformly sampling a batch $B$ of message indices from $I$, secondly computing the batch estimations by applying the full end-to-end model, thirdly computing the categorical cross-entropy\footnote{As mentioned in~\cite{oshea2017introduction}, the sparse categorical cross-entropy between $i$ and $\hat{p}$ is effectively the same loss function. However, with this loss the one-hot encoding layer can be replaced with an \textit{embedding} layer.} of $s_i$ and their corresponding $\hat{p}$, $\forall i \in B$ and lastly performing stochastic gradient descent and backpropagation to update the model weights. This training step is repeated until a stop criterion is reached.

\section{Modified architecture of end-to-end communication system}
\label{sec:proposal}
In training neural networks, hyperparameters such as learning rate and batch size can influence the training speed and resulting accuracy notably. Likewise, the authors of \cite{oshea2017introduction} noted that increasing the batch size during training helps to improve the accuracy of the learned model. As a matter of fact, in the training method outlined above, the batch size influences the normalization layer heavily. For a fixed power constraint $|x_i|^2 = P,~\forall i \in I$, the normalization factor is calculated independently for each batch element; and therefore the effects of the normalization layer are independent of the batch size. For an average power constraint $\E\left[|x_i|^2,~\forall i \in I\right] = P$, this is not the case. In this case, normalization is performed to ensure an average power for the specific batch $B$ with batch size $B_s$:

\begin{equation*}
    \frac{1}{B_s} \sum_{i \in B} |x_i|^2 = P.
\end{equation*}

\begin{figure}[t!]
\centering
\vspace{1mm}
\resizebox{\columnwidth}{!}{

\begin{tikzpicture}
\begin{axis}[
  xlabel = {Batch size},
  ylabel = {Average normalization error},
  ytick={0, 0.05, 0.10, 0.15, 0.20, 0.25},
  xtick={0, 16, 32, 64, 128, 256, 512},
  xticklabels={0, , ,64, 128, 256, 512},
  x label style={font=\footnotesize},
  y label style={font=\footnotesize},
  xticklabel style={font=\footnotesize},
  yticklabel style={font=\footnotesize},
  legend style={nodes={scale=0.75, transform shape}},
  xmin=0,
  xmax=530,
  ymin=0,
  legend entries = {$M=16$, $M=64$, $M=256$},
  legend cell align={left},
  reverse legend,
  scaled y ticks=base 10:2,
  width={\columnwidth},
  height=5.85cm,
  grid=both,
  cycle list={
      {black, mark=*, line width=0.5pt, solid, mark options={solid, line width=0.5pt, fill=white}},
      {blue, mark=*, line width=1pt, dotted, mark options={solid, line width=0.5pt, fill=white}},
      {red, mark=*, line width=0.8pt, dashed, mark options={solid, line width=0.5pt, fill=white}}
  }
]
  \addplot table {figures/normalization_error/data_16.dat};
  \addplot table {figures/normalization_error/data_64.dat};
  \addplot table {figures/normalization_error/data_256.dat};
\end{axis}
\end{tikzpicture}}
\caption{Average normalization error for various batch sizes and number of messages $M$. Experiment performed on randomly initialized transmitter networks, with batches drawn uniformly from $I$, and using a fixed $E_b=1$, i.e. $P=\log_2(M)$, to enable a fair comparison for varying $M$. An average of 30 transmitter network initializations (comprising 2 hidden layers with 60 neurons each) and 1000 uniformly sampled batches is presented.}
\label{fig:normalization-error}
\end{figure}
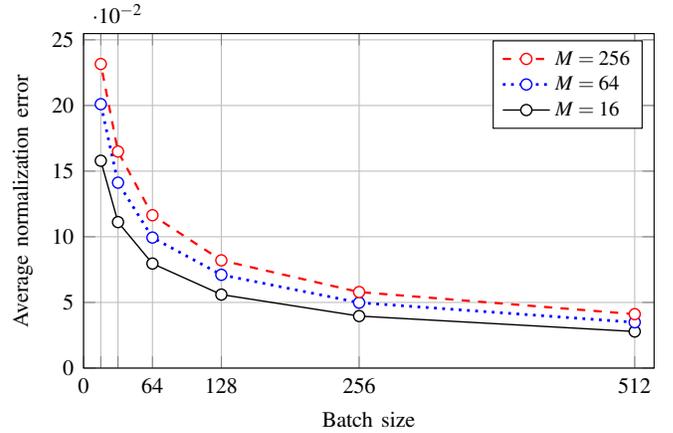

However, for some batches this constraint may not reflect the desired normalization $\E\left[|x_i|^2,~\underline{\forall i \in I}\right] = P$. This effect is visualized in Fig.~\ref{fig:normalization-error}. By picking the batch size too low, in general the batch does not represent $I$ and the average normalization error\footnote{We define the average normalization error of a batch $B$ as $(\sum_{i \in B} \left|x_i - x^{\prime}_i\right|) / B_s$ with $x_i$ and ${x^{\prime}_i}$ the transmitted symbols computed by applying the full transmitter network to respectively $B$ and $I$.} of the batch will be high. For a higher number of messages $M$, the batch size should be increased to have a constant average normalization error.

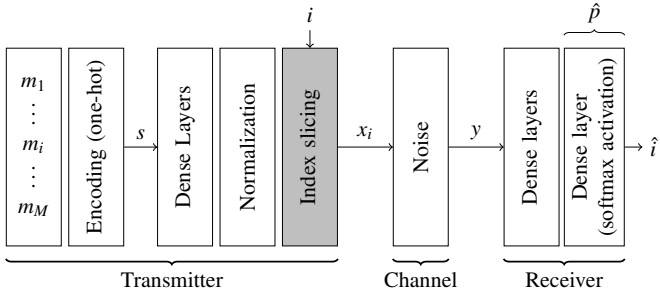
\begin{figure}[t!]
\centering
\resizebox{\columnwidth}{!}{

\begin{tikzpicture}
\tikzstyle{none}=[]
\tikzstyle{hbox}=[fill=white, draw=black, shape=rectangle, minimum height=90pt, minimum width=25pt, align=center]
\tikzstyle{vbox}=[fill=white, draw=black, shape=rectangle, minimum height=25pt, minimum width=90pt, align=center, rotate=90, anchor=north]
\tikzstyle{pointer}=[->]
\tikzstyle{brace}=[thick, decoration={brace, mirror, raise=0.2cm}, decorate]
\tikzstyle{brace text}=[below=8pt]
\tikzstyle{brace up}=[decoration={brace, raise=0.2cm}, decorate]
\tikzstyle{brace up text}=[above=8pt]
\tikzstyle{pointer text}=[above]
		\node (tx1) [style=hbox] {$m_1$ \\ $\vdots$ \\ $m_i$ \\ $\vdots$ \\ $m_M$};
		\node (tx2) [right=3pt of tx1.east, style=vbox] {Encoding (one-hot)};
		\node (tx3) [right=15pt of tx2.south, style=vbox] {Dense Layers};
		\node (tx4) [right=3pt of tx3.south, style=vbox] {Normalization};
		\node (tx5) [right=3pt of tx4.south, style=vbox, fill=lightgray] {Index slicing};
		\node(in) [above=8pt of tx5.east] {$i$};

		\node (ch1) [right=25pt of tx5.south, style=vbox] {Noise};

		\node (rx1) [right=25pt of ch1.south, style=vbox] {Dense layers};
		\node (rx2) [right=2pt of rx1.south, style=vbox] {Dense layer \\ (softmax activation)};

		\node (out) [right=8pt of rx2.south] {$\hat{i}$};

		\draw [style=pointer] (in.south) to (tx5.east);
		\draw [style=pointer] (rx2.south) to (out.west);
		\draw [style=pointer] (tx2.south) to node[pointer text] {$s$} (tx3.north);
		\draw [style=pointer] (tx5.south) to node[pointer text] {$x_i$} (ch1.north);
		\draw [style=pointer] (ch1.south) to node[pointer text] {$y$} (rx1.north);

		\draw [style=brace up] (rx2.north east) to node[brace up text] {$\hat{p}$} (rx2.south east);

		\draw [style=brace] (tx1.south west) to node[brace text] {Transmitter} (tx5.south west);
		\draw [style=brace] (ch1.north west) to node[brace text] {Channel} (ch1.south west);
		\draw [style=brace] (rx1.north west) to node[brace text] {Receiver} (rx2.south west);
\end{tikzpicture}}
\caption{We propose shifting the position of the batch slicing in the end-to-end architecture after the normalization layer.}
\label{fig:arch-proposed}
\end{figure}

A high batch size relative to the number of messages $M$ seems to solve this normalization error issue. However, this is not always convenient, especially for a higher number of messages. To counter normalization errors in another way, we propose changing the architecture of Fig.~\ref{fig:arch-detailed} to the architecture of Fig.~\ref{fig:arch-proposed}. By shifting the position of the batch slicing after the normalization layer, normalization is performed on the messages $m_i,~\forall i \in I$, and hence normalization will ensure the desired constraint of $\E\left[|x_i|^2,~\underline{\forall i \in I}\right] = P$. Therefore, the proposed architecture has no normalization errors and normalization is performed independently of any batch size. The training of this proposed system is also modified slightly: backpropagation should now be computed over a slicing layer. Frameworks such as Tensorflow~\cite{tensorflow2015-whitepaper} (used in our experiments) have built-in support for these kinds of operations.

\section{Experiments}
\label{sec:experiments}
\begin{figure}[t!]
    \centering
    \resizebox{\columnwidth}{!}{

\begin{tikzpicture}
    \begin{groupplot}[
        group style={
            group size=1 by 6,
            xlabels at=edge bottom,
            xticklabels at=edge bottom,
            vertical sep=0pt
        },
        width={\columnwidth},
        height=2.5cm,
        boxplot,
        table/y=sparse_categorical_accuracy,
        xmin=0.8,xmax=1.0005,
        xlabel={Categorical accuracy},
        x label style={font=\small},
        xticklabel style={font=\small},
        xticklabel={\pgfmathparse{\tick*100}\pgfmathprintnumber{\pgfmathresult}\%},
        ytick={1.5},
        yticklabels={\empty},
        yticklabel style={font=\small},
        xtick pos=left,
        ytick pos=left,
        cycle list={
            {red, solid, mark=o, line width=0.6pt, mark options={solid, line width=0.6pt, fill=white}, pattern=crosshatch dots, pattern color=red},
            {blue, solid, mark=o, line width=0.6pt, mark options={solid, line width=0.6pt, fill=white}},
        },
        area legend,
    ]
    \nextgroupplot[yticklabels={$B_s=16$}, legend to name=legend, legend columns=2, reverse legend, legend style={nodes={scale=0.85, transform shape}}]
        \coordinate (top) at (rel axis cs:0.5,1);
        \addlegendentry{Existing};
        \addlegendentry{Modified};
        \addplot table {figures/comparison/data_BEFORE_16.dat};
        \addplot table {figures/comparison/data_AFTER_16.dat};
    \nextgroupplot[yticklabels={$B_s=32$}]
        \addplot table {figures/comparison/data_BEFORE_32.dat};
        \addplot table {figures/comparison/data_AFTER_32.dat};
    \nextgroupplot[yticklabels={$B_s=64$}]
        \addplot table {figures/comparison/data_BEFORE_64.dat};
        \addplot table {figures/comparison/data_AFTER_64.dat};
    \nextgroupplot[yticklabels={$B_s=128$}]
        \addplot table {figures/comparison/data_BEFORE_128.dat};
        \addplot table {figures/comparison/data_AFTER_128.dat};
    \nextgroupplot[yticklabels={$B_s=256$}]
        \addplot table {figures/comparison/data_BEFORE_256.dat};
        \addplot table {figures/comparison/data_AFTER_256.dat};
    \nextgroupplot[yticklabels={$B_s=512$}]
        \coordinate (legend) at (rel axis cs:0, 0.5);
        \addplot table {figures/comparison/data_BEFORE_512.dat};
        \addplot table {figures/comparison/data_AFTER_512.dat};
    \end{groupplot}
    \node[right=0pt] at (legend) {\pgfplotslegendfromname{legend}};
\end{tikzpicture}}
    \caption{Comparison of validation accuracy (averaged over 30 batches) after training both the existing architecture of Fig.~\ref{fig:arch-detailed} and our modified architecture of Fig.~\ref{fig:arch-proposed}. Training is performed for various batch sizes $B_s$ and a fixed $M=128$, $SNR=45$~dB, 76.800 data points (split in batches), with Adam optimizer with learning rate 0.008 and receiver and transmitter networks both comprising 2 dense layers with 100 neurons each. To have representative results, training is performed for 10 random network initializations and 10 data generators.}
    \label{fig:comparison-accuracy}
\end{figure}
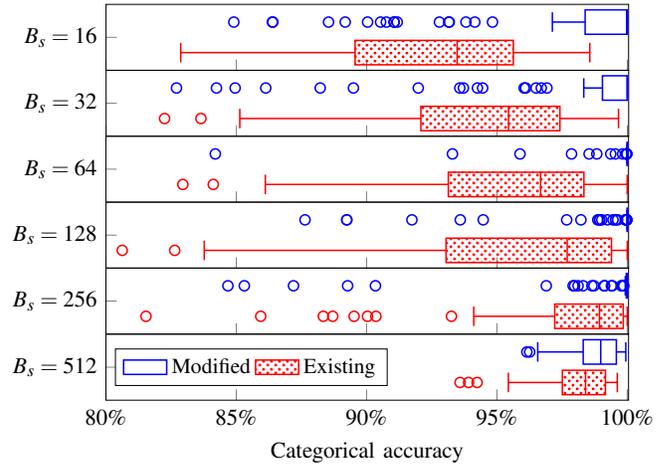

The aim of this section is to experimentally determine the influence of our modified architecture on the training and the performance of the model. To this end, we train both architectures on the same batches, while having the same network initializations. At the end of training, the categorical accuracy of the networks is determined. To only measure the effect of normalization errors while training, the validation set is chosen such that the normalization error is 0. Fig.~\ref{fig:comparison-accuracy} shows the results for a fixed $M=128$ and $SNR=45$~dB, but for other configurations we noticed similar results. Note: for $B_s \in \{16,\ldots,256\}$, the median of the categorical accuracy for the modified architecture is 100\%. Also, due to the low error, only outliers can be seen for $B_s=\{64, 128, 256\}$.

We make three main observations. Firstly, for each architecture, training on a higher batch size $B_s$ (up to $B_s=256$) on average leads to a better performing model. This is expected, since a batch size which is too small usually leads to an unstable learning curve. For a batch size $B_s=512$, the accuracy drops slightly. In this case, we inspected the learning curve and noticed that it did not yet converge; which is explained by the fact that at a higher batch size $B_s$, the number of gradient updates are lower for a fixed number of data points. For lower batch sizes, the learning curves have converged: either to a stable constant 100\% accuracy or oscillating around a sub-optimal solution with less than 100\% accuracy.

Secondly, the accuracy of the resulting model increases significantly with our modified architecture. Normalization errors clearly have a negative impact on the training of end-to-end communication systems. This is clarified as follows: in the existing architecture, the receiver should not only learn to cope with an additive noise channel, but also with the introduced normalization errors. Therefore, the extra normalization errors make the learning task harder for the receiver.

Thirdly, with an increase in batch size $B_s$, the accuracy difference induced by our proposed modification diminishes. This is explained by the decline in normalization error with a growing batch size (see Fig.~\ref{fig:normalization-error}). Nevertheless, the accuracy difference remains significant, and therefore, using our proposed modification scales better in terms of batch size. This is especially important for a higher number of messages, which require a higher batch size to reduce the normalization errors.

\section{Conclusion}
\label{sec:conclusion}
In this paper, we have focused on the architecture of an end-to-end communication system and have proposed shifting the position of the batch slicing after the normalization layer. The advantage of the proposed modification is that it does not incur normalization errors when training the end-to-end system. We showed that normalization errors make the learning task of the receiver harder. Therefore, in general using our proposed modification leads to significantly improved performance of trained models than existing architectures (trained for similar learning configurations). Furthermore, the training of our modified architecture scales much better in terms of the batch size: even at batch sizes lower than the number of messages, performance remains competitive.

\section*{Acknowledgment}
This research has received funding from the European Union's Horizon 2020 research and innovation programme under Grant Agreement No. 101017171 (MARSAL project).

\bibliographystyle{IEEEtran}
\bibliography{main}

\end{document}